\documentclass[twocolumn,showpacs,preprintnumbers,amsmath,amssymb]{revtex4}

\usepackage{graphicx}
\usepackage{dcolumn}
\usepackage{bm}

\begin{document}

\preprint{Submitted July, 2008}

\newcommand{\comm}[1]{\textcolor{blue}{(#1)}}

\title{Fluctuations and Pseudo Long Range
Dependence in Network Flows: A Non-Stationary Poisson Process Model}

\author{Yudong Chen, Li Li, Yi Zhang, Jianming Hu}
 \altaffiliation{Department of Automation, Tsinghua University, Beijing 100084, P. R. China}
 \email{li-li@mail.tsinghua.edu.cn}

\date{\today}

\begin{abstract}
In the study of complex networks (systems), the scaling phenomenon
of flow fluctuations refers to a certain power-law between the mean
flux (activity) $\langle F_i \rangle$ of the $i$th node and its
variance $\sigma_i$ as $\sigma_i \propto \langle F_{i} \rangle
^{\alpha}$. Such scaling laws are found to be prevalent both in
natural and man-made network systems, but our understanding of their
origins still remains limited. In this paper, a non-stationary
Poisson process model is proposed to give an analytical explanation
of the non-universal scaling phenomenon: the exponent $\alpha$
varies between $1/2$ and $1$ depending on the size of sampling time
window and the relative strength of the external/internal driven
forces of the systems. The crossover behavior and the relation of
fluctuation scaling with pseudo long range dependence are also
accounted for by the model. Numerical experiments show that the
proposed model can recover the multi-scaling phenomenon.
\end{abstract}

\maketitle

\section{Introduction}
\label{sec:intro}

The studies of the complex networks have attracted increasing
interests since the last decade
\textsuperscript{\cite{AlbertBarabasi2002}, \cite{Newman2003},
\cite{BoccalettiaLatorabMorenodChavezfHwanga2006},
\cite{GaoSunWu2007}}. Much work has been devoted to the
understanding of the dynamic processes of network flows and
significant advances has been achieved. However, there are still
many questions which remain to be answered thoroughly. Among them
the origin of the fluctuations of network flow is of particular
interest, which will be addressed in this paper.

The study of the scaling of fluctuations can be dated back to the
Taylor's law initialized in \cite{Taylor1961}. Recently, related
studies attract increasing interests again, since Menezes and
Barabasi's proposal of the so called scaling analysis. The recent
research aims to describe the dynamics of a large number of nodes
simultaneously and examine the collective behaviors of the systems
\textsuperscript{\cite{MenezesBarabasi2004a},
\cite{MenezesBarabasi2004b}, \cite{ EislerKerteszYookBarabasi2006},
\cite{EislerBartosKertesz2008}}. They investigated the coupling
between the average flux and fluctuations and found that in complex
networks there exists a characteristic coupling between the mean and
variance of flux on individual nodes as $\sigma_i \propto \langle
F_{i} \rangle ^ {\alpha}$. Such scaling law was found to hold in
diverse natural and man-made systems. Moreover, they claimed that
real systems belong to one of two distinct universality classes. The
first class yields $\alpha \approx 1/2$, which indicates that system
dynamics are dominated by internal factors. Internet and microchip
systems belong to this class. The other class yields $\alpha \approx
1$, which indicates that the fluctuations mainly come from external
driven forces. Highway traffic, river networks, and the World Wide
Web belong to this class.

These interesting findings are soon followed by further studies. The
scaling phenomenon has been confirmed in a wide range of complex
networks, e.g. stock markets
\textsuperscript{\cite{EislerKertesz2006},
\cite{RaffaelliMarsili2006}}, gene systems
\textsuperscript{\cite{NacherOchiaiAkutsu2005}}, download networks
\textsuperscript{\cite{HanLiuMa2008}}, software developments
\textsuperscript{\cite{Valverde2007}}, and urban transportation
networks \textsuperscript{\cite{ChenLiZhangHuJin2008}}. In these
systems, a rich variety of exponent $\alpha$ from $1/2$ to $1$ can
be observed, which approach $1$ as the size of sampling time window
or the relative strength of the external driven forces to the
internal dynamics increases. Some systems even exhibit the so called
multi-scaling behavior (i.e., the power-law extends to the $q$th
moments of $F_i$) and the crossover (i.e., different groups/types of
nodes in the same system possess different values of $\alpha$)
behavior \textsuperscript{\cite{EislerKertesz2006},
\cite{RaffaelliMarsili2006}, \cite{Valverde2007},
\cite{KujawskiTadicRodgers2007}}. These phenomena are believed to
reflect the competition between the external and internal dynamics,
as well as the inhomogeneity of the systems' structure
\textsuperscript{\cite{RaffaelliMarsili2006},
\cite{ChenLiZhangHuJin2008}, \cite{EislerKertesz2005}}.

Several models have been proposed to explain the underlying
mechanisms that generate the scaling laws
\textsuperscript{\cite{MenezesBarabasi2004b},
\cite{EislerKerteszYookBarabasi2006},
\cite{KujawskiTadicRodgers2007}, \cite{EislerKertesz2005},
\cite{DuchArenas2006}, \cite{DuchArenas2007}}, most of which are
based on random walk or diffusive dynamics models. These models can
reproduce the scaling phenomenon and link the dynamic processes of
networks with the associated network topologies. Usually, the
following questions are addressed:

\begin{enumerate}

\item The scaling law is better to be represented via both explicit analytical
expression and numerical simulations
\textsuperscript{\cite{EislerBartosKertesz2008},
\cite{EislerKertesz2005}, \cite{DuchArenas2006},
\cite{DuchArenas2007}, \cite{ChenLiZhangJin2008},
\cite{MeloniGardenesLatoraMoreno2008}}.

\item The non-universality values of $\alpha$
from $1/2$ to $1$, the effect of the sampling time windows and the
internal/external driven forces should be explained in an integrated
framework \textsuperscript{\cite{EislerBartosKertesz2008},
\cite{ChenLiZhangHuJin2008}, \cite{MeloniGardenesLatoraMoreno2008}}.

\item The crossover phenomenon should be accounted for
\textsuperscript{\cite{EislerBartosKertesz2008},
\cite{ChenLiZhangJin2008} \cite{MeloniGardenesLatoraMoreno2008}}.

\end{enumerate}

However, we are still interested in finding a even more simpler
model other than random walk/diffusive dynamics models to answer the
above questions. In the authors' earlier attempt, a non-stationary
Poisson process model was proposed in \cite{ChenLiZhangJin2008} to
overcome the above shortcomings. This model could explain, both
analytically and numerically, the transition of $\alpha$ from 1/2 to
1, as well as its dependence on the sampling time windows and
external/internal driven forces. A similar model also appeared in
\cite{MeloniGardenesLatoraMoreno2008}. However, both
\cite{ChenLiZhangJin2008} and \cite{MeloniGardenesLatoraMoreno2008}
assumed that the length of the intervals between two consecutive
jumps of the arriving rate equals that of the sampling time window.
In order to better model the phenomena in real systems, it is
necessary to relax this restriction and allow different size of the
jumping intervals and the sampling time windows.

This paper extends the model in \cite{ChenLiZhangJin2008} to more
general cases and provides a more comprehensive study. In
particular, analytical solutions are provided for $\alpha$'s
dependence on sampling time windows and external/internal driven
forces, as well as for the crossover phenomenon. Moreover, the
connection of fluctuation scaling with pseudo long range dependence
is explained. The multi-scaling behaviors are also investigated
numerically. The proposed model is able to capture the essential
mechanism that drives the fluctuations of flow and offer a unified
and concise picture for the various phenomena observed in real
traffic flows.

\section{The Non-Stationary Poisson Process Model}
\label{sec:model}

Suppose the network system consists of $N$ nodes, which will be
indexed as $i=1$, ..., $N$ in the rest of this paper. The arriving
flow $f_i(t)$ of the $i$th node at the $t$th time interval is
assumed to behave as a non-stationary Poisson process
\begin{equation}
\label{eq.1}
  \Pr (f_i (t) = n) = \frac{e^{-\lambda_i (t)} {\lambda_i (t)}^n}{n!}
\end{equation}

\noindent where $n=1$, $2$, ... . $\Pr$ denotes the abbreviation of
the probability.

Moreover, the varying process of the arriving rate $\lambda_i (t)$
is assumed to change according to a network-wide finite-state
deterministic or stochastic process. The system will enter one state
during a certain time interval $\tau$ and then enter another state,
which as a result changes the arriving rate as well as the arriving
flux of the Poisson process.

For instance, the widely used Markov-modulated Poisson process
(MMPP) model is a typical example of such models. It is a doubly
stochastic process where the intensity of a Poisson process is
defined by the state of a Markov chain. The Markov chain can
therefore be said to modulate the Poisson process, and thus comes
the name MMPP in many literatures
\textsuperscript{\cite{Rossiter1987},
\cite{FischerMeier-Hellstern1992}, \cite{AndersenNielsen1998}
  \cite{SalvadorValadasPacheco2003}
  \cite{MuscarielloMelliaMeoAjmoneMarsanLoCigno2005},
\cite{BolchGreinerdeMeerTrivedideMeerTrivedi2006}}. Fig.~1 shows a
Poisson process modulated by a two-state Markov process, in which we
can observe the shift of average arriving flow every $\tau$.

\vskip 2mm

\centerline{\includegraphics[width=3in]{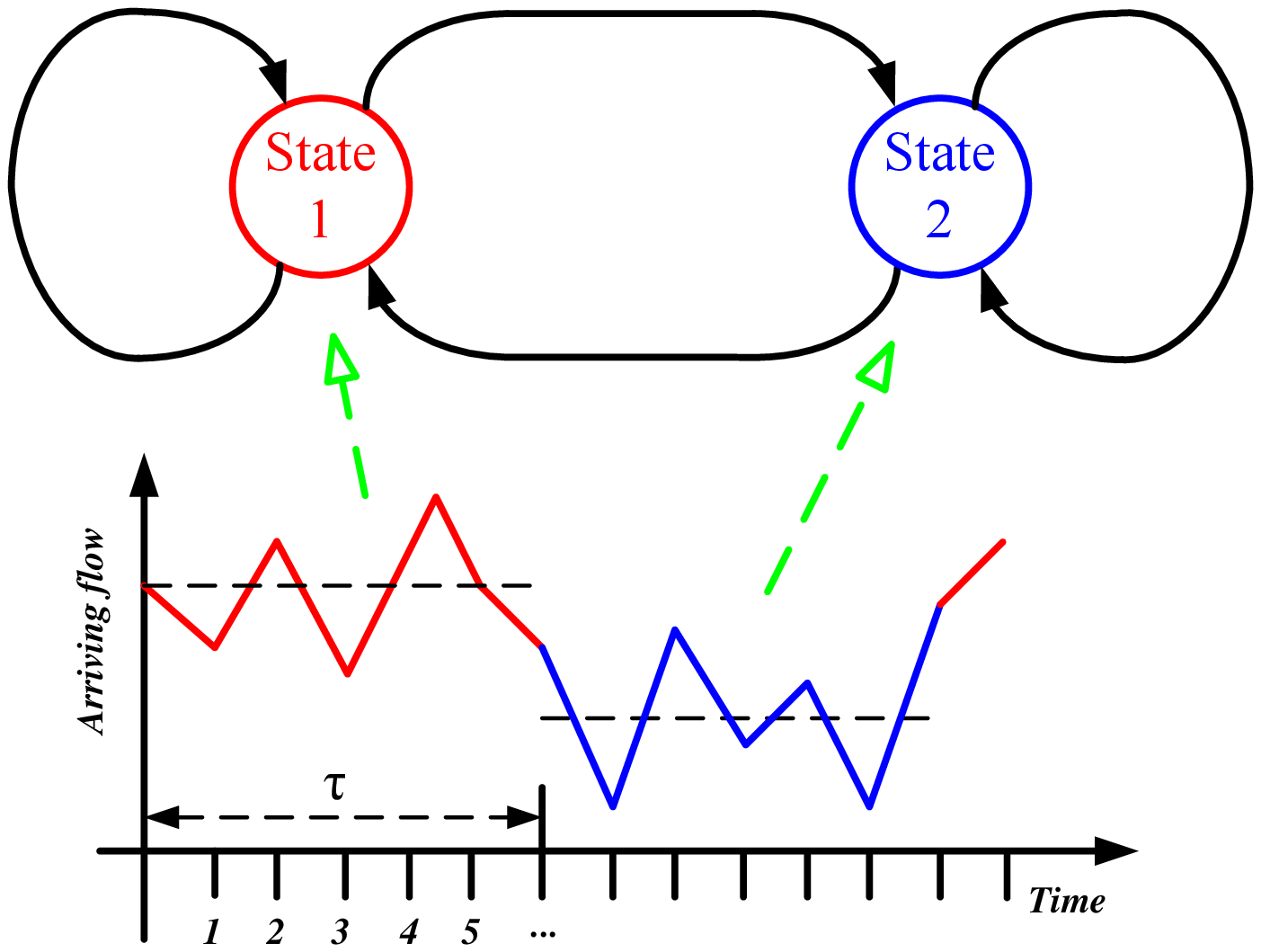}}
\centerline{\footnotesize
\begin{tabular}{p{7.5cm}} \bf Fig.\,1. \rm Examples of a Poisson process modulated by a two-state
Markov process.\end{tabular}}

\vskip 2mm

Before discussing the characteristics of such models, let's
introduce some phrases to distinguish the different time intervals
employed in the rest of this paper.

\begin{enumerate}

\item Intrinsic time interval denotes the minimum discernable (from the
viewpoint of observer) time interval in the model, which is
determined only by the non-stationary Poisson process. And we use
the integer number set $\{t = 1, 2, 3, ..., T\}$ to index the
intrinsic time points.

\item Endogenous jump time interval denotes the lasting time after the
process enters one state and before it leaves it, which is
determined only by the non-stationary Poisson process, too. Here, we
assume it to be an integer multiple of the intrinsic time interval
and use the integer number set $\{j = 1, 2, 3, ..., T/\tau \}$ to
index the jump time points.

\item Exogenous sampling time interval denotes the length of the
sampling window (from the viewpoint of observer), which can be
changed by the observer. We assume it to be an integer multiple of
the intrinsic time interval, too. We use the symbol $\Delta t$ to
represent it and the integer number set $\{s = 1, 2, 3, ...,
T/\Delta t \}$ to index the sampling time points.

\end{enumerate}

Suppose the dynamic varying process of the arriving rate $\lambda_i
(t)$ can be depicted as
\begin{equation}
\label{eq.2}
  \lambda_i(t) = k_i \lambda(j), \indent
  \textrm{for} \indent (j-1) \tau + 1 \leq t \leq j \tau
\end{equation}

\noindent where $k_i$ is a scale coefficient which accounts for the
factors controlling the relative magnitude of flow on node $i$. For
example, in a random walkers model, $k_i$ stands for the degree of
node $i$ \textsuperscript{\cite{MeloniGardenesLatoraMoreno2008}}.

$\lambda(j)$ represents the network-wide stochastic scalar parameter
at the $j$th jumping interval. We assume all the nodes in the
network systems follow a synchronized transition tempo so as to
reproduce the network-wide fluctuations in a simple way. Studies
show that we can relax this assumption to reproduce even more
complex phenomena.

In this paper, we mainly focus on the steadily state of this
process, since it determines the steady distribution of the scalar
parameter $\lambda(j)$ and thus shapes the mean and variance of flux
$f_i (t)$ at the $i$th node. Suppose we have altogether $M$ possible
values of the scalar parameter $\lambda(j)$, which are denoted as
$\{\lambda_m | m=1, ..., M\}$; and each $\lambda_m$ corresponds to a
certain state of this process. Moreover, we assume $\lambda(j)$ have
the following stationary distribution
\begin{equation}
\label{eq.3}
  \Pr(\lambda(j) = \lambda_m) = p_m
\end{equation}

In practice, we usually deal with sampled flows series instead of
the original series $f_i(t)$. Given a sampling window of length
$\Delta t$, the observed flow data at the $s$th sampling index is
the summation of the original flow values within the sampling
window:
\begin{equation}
\label{eq.4}
 F_i^{\Delta t} (s) = \sum_{t=(s-1)\Delta t + 1}^{s \Delta t} f_i(t), \indent
 \textrm{for} \indent s = 1,..., T/\Delta t
\end{equation}

Under the assumption of ergodicity, the time average $\langle \cdot
\rangle$ of the flow series equals its equilibrium value. Thus the
observed average flow at node $i$ can be written as
\begin{eqnarray}
\label{eq.5}
  & & \langle F_i^{\Delta t} \rangle \nonumber \\
  & = & \frac{1}{T/\Delta t} \sum_{s=1}^{T/\Delta t} F_i^{\Delta t} (s) 
  = \frac{\Delta t}{T} \sum_{s=1}^{T/ \Delta t} \sum_{t=(s-1)\Delta t+1}^{s \Delta t} f_i(t) \nonumber \\
  & = & \frac{\Delta t}{T} \sum_{t=1}^{T} f_i(t) 
  = \frac{\Delta t}{T} \sum_{j=1}^{T/\tau} \sum_{t=(j-1)\tau + 1}^{j \tau} f_i(t) \nonumber \\
  & = & \frac{\Delta t}{T} \sum_{j=1}^{T/\tau} k_i \lambda(j) \tau 
  = k_i \Delta t \frac{1}{T/ \tau} \sum_{j=1}^{T/\tau} \lambda(j) \nonumber \\
  & = & k_i \Delta t \langle \lambda \rangle
\end{eqnarray}

\noindent where $\langle \lambda \rangle := \frac{1}{T/ \tau}
\sum_{j=1}^{T/\tau} \lambda(j) = \sum_{m=1}^{M} \lambda_m p_m$.

To derive the variance $(\sigma_i^{\Delta t})^2$ of the observed
flow, we need to distinguish two cases in terms of sampling window
size $\Delta t$.

\vskip 2mm

\centerline{\includegraphics[width=3in]{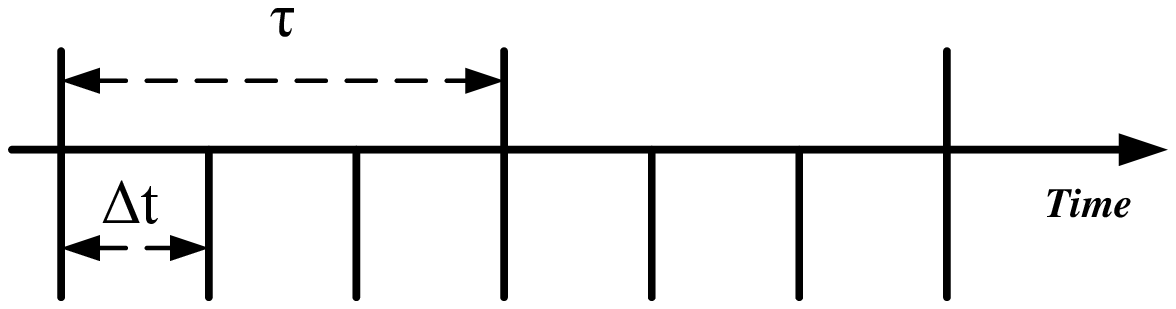}}
\centerline{\footnotesize
\begin{tabular}{p{7.5cm}} \bf Fig.\,2. \rm Diagram of Case I), where $\Delta t \leq \tau$.\end{tabular}}

\vskip 2mm

I) $\Delta t \leq \tau$ (see Fig.~2). Further assume that $\tau =
\kappa \Delta t$, $\kappa \in \mathbb{N} $, thus the arriving rate
keeps constant in each sampling window. Noting that the $s$th
observed data satisfies $F_i^{\Delta t} (s) \sim P(k_i \lambda_j
\Delta t)$ for $(j-1) \kappa + 1 \leq s \leq j \kappa$, we have
\begin{eqnarray}
\label{eq.6}
  & & (\sigma_i^{\Delta t})^2 \nonumber \\
  & = & \frac{1}{T/ \Delta t}
            \sum_{s=1}^{T/ \Delta t} \left[ F_i^{\Delta t} (s) - \langle F_i^{\Delta t}\rangle \right]^2 \nonumber \\
  & = & \frac{\Delta t}{T}
            \sum_{j=1}^{T / \tau}
            \sum_{s = (j-1) \kappa + 1}^{j \kappa}
                \left[ F_i^{\Delta t}(s) \right]^2
        - \langle F_i^{\Delta t}\rangle ^2 \nonumber \\
  & = & \frac{\Delta t}{T}
            \sum_{j=1}^{T/ \tau} \kappa \left[\frac{1}{\kappa}
            \sum_{s=(j-1) \kappa + 1}^{j \kappa} \left[ F_i^{\Delta t}(s) \right]^2 \right]
        - \left( k_i \Delta t \langle \lambda \rangle \right) ^2 \nonumber \\
  & = & \frac{\Delta t}{T}
            \sum_{j=1}^{T/ \tau} \kappa \left[ k_i \lambda(j) \Delta t + (k_i \lambda(j) \Delta t)^2 \right]
        - \left( k_i \Delta t \langle \lambda \rangle \right) ^2 \nonumber \\
  & = & k_i \Delta t \langle \lambda \rangle
        + k_i^2 (\Delta t)^2 \left[ \textrm{Var} \lambda + \langle \lambda \rangle ^2 \right]
        - \left( k_i \Delta t \langle \lambda \rangle \right) ^2 \nonumber \\
  & = & k_i \Delta t \langle \lambda \rangle
        + k_i^2 (\Delta t)^2 \textrm{Var} \lambda \nonumber \\
  & = & \langle F_i^{\Delta t} \rangle
        + \frac{\textrm{Var} \lambda}{\langle \lambda \rangle ^2} \langle F_i^{\Delta t} \rangle ^2
\end{eqnarray}

\noindent where $\textrm{Var} \lambda: = \frac{\tau}{T}
\sum_{j=1}^{T/ \tau} \left[\lambda(j) - \langle \lambda
\rangle\right]^2 = \sum_{m=1}^M \lambda_m^2 p_m - \left[\sum_{m=1}^M
\lambda_m p_m\right]^2$.

\vskip 2mm

\centerline{\includegraphics[width=3in]{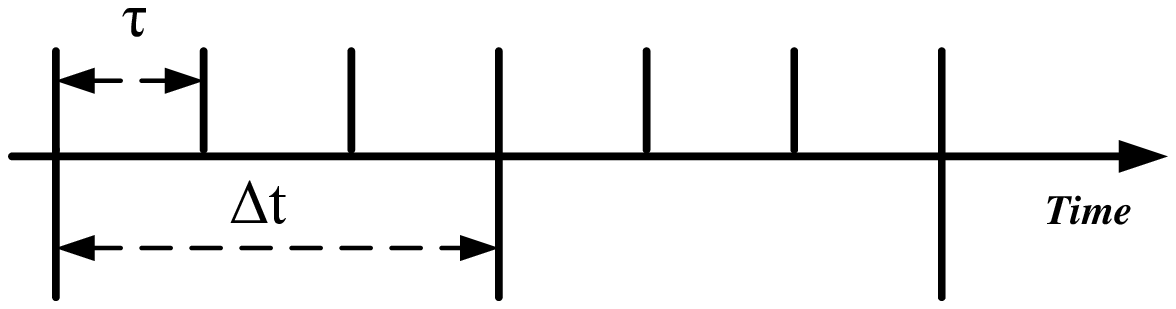}}
\centerline{\footnotesize
\begin{tabular}{p{7.5cm}} \bf Fig.\,3. \rm Diagram of Case II), where $\Delta t > \tau$.\end{tabular}}

\vskip 2mm

II) $\Delta t > \tau$ (see Fig.~3). Further assume that $\Delta t =
\kappa \tau$, $\kappa \in \mathbb{N} \setminus \{ 1 \}$. We have
\begin{eqnarray}
\label{eq.7}
  & & (\sigma_i^{\Delta t})^2 \nonumber \\
  & = & \frac{\Delta t}{T}
            \sum_{s=1}^{T/ \Delta t} [ F_i^{\Delta t}(s) ] ^2
        - \langle F_i^{\Delta t}\rangle ^2 \nonumber \\
  & = & \frac{\Delta t}{T} \sum_{s=1}^{T/ \Delta t} \left[
            \sum_{j=(s-1)\kappa+1}^{s \kappa}
                \sum_{t=(j-1)\tau+1}^{j\tau} f_i(t) \right] ^2
        - \langle F_i^{\Delta t}\rangle ^2 \nonumber \\
\end{eqnarray}

Let $\Lambda_s := \frac{1}{\kappa} \sum_{j=(s-1) \kappa + 1}^{s
\kappa} \lambda(j) $ denote the average of $\lambda(j)$ within the
$s$th sampling window. Noting that $\langle \Lambda_s \rangle =
\frac{\Delta t}{T} \sum_{s=1}^{T/ \Delta t} \Lambda_s = \langle
\lambda \rangle$, Eq.(\ref{eq.7}) can be simplified as
\begin{eqnarray}
\label{eq.8}
  & & (\sigma_i^{\Delta t})^2 \nonumber \\
  & = & k_i \Delta t \langle \Lambda_s \rangle
        + k_i^2 (\Delta t)^2 \left[ \kappa \textrm{Var} \Lambda_s + \langle \Lambda_s \rangle ^2 \right]
        - \left( k_i \Delta t \langle \Lambda_s \rangle \right) ^2 \nonumber \\
  & = & \langle F_i^{\Delta t} \rangle
        + \frac{\kappa \textrm{Var} \Lambda_s}{\langle \Lambda_s \rangle ^2} \langle F_i^{\Delta t} \rangle ^2
\end{eqnarray}

Moreover, we have
\begin{equation}
\label{eq.9}
  \textrm{Var} \Lambda_s = \frac{1}{T/\Delta t}
                \sum_{s=1}^{T/\Delta t} \left(\Lambda_s - \langle \Lambda_s \rangle
                \right)^2 = \frac{1}{\kappa} \textrm{Var} \lambda
\end{equation}

Thus, we can summarize Eq.(\ref{eq.5}), (\ref{eq.6}) and
(\ref{eq.8}) as
\begin{equation}
\label{eq.10}
    \langle F_{i}^{\Delta t} \rangle
    = k_i \Delta t \langle \lambda \rangle
\end{equation}

and
\begin{equation}
\label{eq.11}
  (\sigma_i^{\Delta t})^2 = k_i \Delta t \langle \lambda \rangle
        + (k_i \Delta t)^2 \textrm{Var} \lambda \\
    = \langle F_i^{\Delta t} \rangle
        + \frac{\textrm{Var} \lambda}{\langle \lambda \rangle^2} \langle F_i^{\Delta t} \rangle^2
\end{equation}

Eq.(\ref{eq.11}) indicates that no matter the comparative ratio
between the size of sampling window and the size of the jump
interval, the variance $(\sigma_i^{\Delta t})^2$ is always a
compound of the mean $F_i^{\Delta t}$ and its power of 2. Therefore,
we can always find the scaling exponent $\alpha$ varies between
$1/2$ and $1$.

Besides, it should be pointed out that although we assume $\kappa$
to take integer values, a non-integer $\kappa$ does not affect the
main conclusions of the proposed model.

\section{The Scaling of Fluctuations and Pseudo Long Range Dependence Reproduced by the Proposed Model}
\label{sec:scaling}

Based on Eq.(\ref{eq.10}) and (\ref{eq.11}), the scaling law of the
proposed model can be analytically explained. Particularly, we will
discuss the three scaling phenomena which have been frequently
observed both in simulations and real systems in previous
literature. The connection to pseudo long range dependence will also
be investigated.

1) The effect of the external and internal forces

\label{sec:3.1}

The time-variation of $\lambda(j)$ can be regarded as the result of
external driven force; thus the ratio $\textrm{Var} \lambda /
\langle \lambda \rangle$ reflects its relative strength to internal
dynamic factors. If the external driven force is weak, we have
$\textrm{Var} \lambda / \langle \lambda \rangle \ll 1/(k_i\Delta
t)$; thus the second term in the r.h.s. of (\ref{eq.11}) can be
omitted, yielding $\alpha = 1/2$. On the other hand, if the external
driven force is strong and $\textrm{Var} \lambda / \langle \lambda
\rangle \gg 1/(k_i\Delta t)$, the first term can be omitted,
yielding $\alpha = 1$. For intermediate values of $\textrm{Var}
\lambda / \langle \lambda \rangle $, the scaling law actually breaks
down \textsuperscript{\cite{EislerKertesz2006a}}, but we still have
effective values of $\alpha$ ranging between $1/2$ and $1$.

Particularly, we are interested in network systems containing two or
more driven forces (some are global/system-wide and external; some
are local/district-wide). To separate these might help
distinguishing which effect is dominant at a certain time, which
will further assist us to take appropriate actions/controls. For
example, in a previous report
\textsuperscript{\cite{ChenLiZhangHuJin2008}}, we found the scaling
phenomena in urban ground traffic network might be explained as the
interaction between the nodes' internal dynamics (i.e. queuing at
intersections, car-following in driving) and the changes of the
external (network-wide) traffic demand (i.e. the every day increase
of traffic amount during peak hours and shocking caused by traffic
accidents). This may allow us to further understand the mechanisms
governing the transportation system's collective behavior.

2) The influence of the sampling window

If the size of the sampling window $\Delta t \ll \langle \lambda
\rangle / (k_i \textrm{Var} \lambda)$, we can omit the second term
in the right-hand side of (\ref{eq.11}), which contain the quadratic
form of $\Delta t$, and thus $\alpha = 1/2$ \footnote{As pointed out
in \textsuperscript{\cite{DuchArenas2006}}, in the limit of
infinitesimal sampling window such that $\langle F_i^{\Delta
t}\rangle \ll 1$, $\alpha = 1/2$ can be always recovered regardless
of the underlying arriving process.}. If $\Delta t \gg \langle
\lambda \rangle / (k_i \textrm{Var} \lambda)$, we can omit the first
term in the right-hand side of (\ref{eq.11}) and thus $\alpha = 1$.
For intermediate values of $\Delta t$, the effective $\alpha$ lies
between $1/2$ and $1$.

3) The crossover behavior

For nodes with small $k_i$, the second term in the right-hand side
of (\ref{eq.11}) can be omitted and it yields $\alpha = 1/2$. For
nodes with large $k_i$, the first term in the right-hand side of
(\ref{eq.11}) can be omitted and thus $\alpha = 1$. If the system
has heterogeneous structure and  consists of a broad range of $k_i$
across the nodes, it can be expected that different subsystems
(i.e., different groups of nodes with different $k_i$ and thus
different $\langle F_i^{\Delta t}\rangle$) have different values of
exponent $\alpha$.

4) The coexistence with pseudo long range dependence

Besides the scaling law of fluctuations, the long range dependence
represents another kind of power-law, which couples the variance and
the size of the sampling window.
\begin{equation}
\label{eq.12}
  \sigma_i (\Delta t) \propto \Delta t ^ {H(i)}
\end{equation}

\noindent where $H(i)$ denotes the well-known Hurst exponent.

The coexistence of these two power-laws has received considerable
interest, since they describe the behavior of the same standard
deviation $\sigma_i(\Delta t)$. Previous studies showed that these
two power-laws coexisted in some systems (e.g., stock markets), and
several models were developed to explain such phenomenon
\textsuperscript{\cite{EislerKertesz2005},
\cite{EislerKertesz2006a}, \cite{LiuMaRenShanZhang2004}}.

In the proposed model, the power-law in (\ref{eq.12}) can be derived
directly from (\ref{eq.12}), with $H(i)$ ranging between $1/2$ and
$1$, which recovers the possible coexistence of long range
dependence (actually ``pseudo long range dependence'' here) and the
scaling of fluctuations. Moreover, (\ref{eq.11}) predicts that $H(i)
\rightarrow 1/2$ when $\textrm{Var} \lambda / \langle \lambda
\rangle \rightarrow 0$, which reflects the uncorrelated nature of a
simple Poisson process. On the other hand,when $\textrm{Var}\lambda
/ \langle \lambda \rangle > 0$, the time-variation of the arriving
rate induces long-range dependency to the modulated Poisson process
and results in $H(i)>1/2$.

It should be pointed out that the long-range dependency discussed
above is indeed the so called ``pseudo'' long range dependency
(PLRD) instead of mathematically rigorous LRD. Usually, LRD means
that the decay of the autocorrelation function is hyperbolic and
decays slower than exponentially for all the time lag (time scale).
However, long range dependency properties, heavy tail distributions,
and all other characteristics of real time series (i.e. Internet
traffic, ground vehicular traffic flow) are meaningful only over a
limited range of time scales. Many recent approaches
\textsuperscript{\cite{AndersenNielsen1998}
  \cite{SalvadorValadasPacheco2003}
  \cite{MuscarielloMelliaMeoAjmoneMarsanLoCigno2005}} had illustrated that
appropriately constructed Markov models appear to be a viable
modeling tool in the context of modeling LRD time series over
several time scales. Particularly, the MMPP models are shown to
provide good matches of LRD properties under large time scales. The
key idea behind is obtaining a model that has fast transients on a
local scale and evolves with a slower time constant between
different groups of states. In other words a model that has some
implicit form of memory, which is the origins of such PLRD. The
model proposed here yields some kind of PLRD also via this way.

Besides, the proposed model also indicates clear non-universality,
which means that the Hurst exponents of different nodes may vary
even though the underlying Poisson mechanisms for these nodes are
the same. Thus we should be careful when applying concepts like
scaling and universality to complex systems
\textsuperscript{\cite{EislerKertesz2006b}}.

\section{Numerical Experiments}
\label{sec:experiment}

In this section, some numerical experiments are designed to verify
the proposed model.


\vskip 2mm

\centerline{\includegraphics[width=3in]{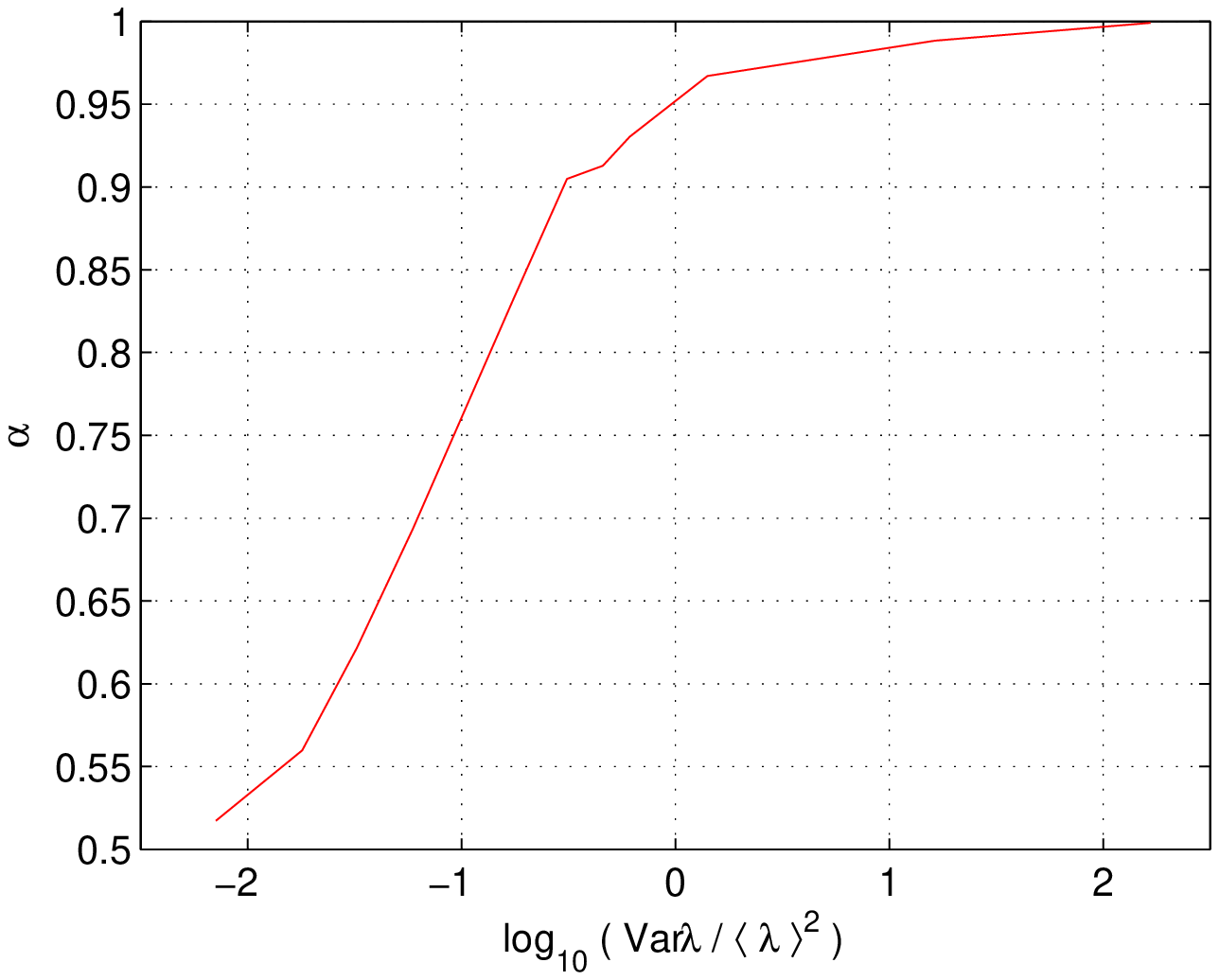}}
\centerline{\footnotesize
\begin{tabular}{p{7.5cm}} \bf Fig.\,4. \rm The transition behavior of $\alpha$ w.r.t the
  competition of the external/internal driven forces produced
  by simulation.\end{tabular}}

\vskip 2mm

In order to observe the influence of the strength of the external
and internal forces, the corresponding parameters of the model are
set as $T=100000$, $\tau=1$, $\Delta t =1$. $N=20$, $k = [1, 2, 3,
..., 20]$. $M=10$, the ratio $Var\lambda / \langle \lambda \rangle$
is controlled to vary from $0.01$ to $200$. In the first test. The
transition of $\alpha$ from $1/2$ to $1$ under different
$\textrm{Var} \lambda / \langle \lambda \rangle$ can be clearly
observed from Fig.~4. Similar results can be found that $\Delta t >
\tau$ cases and are thus omitted here.


\vskip 2mm

\centerline{\includegraphics[width=3in]{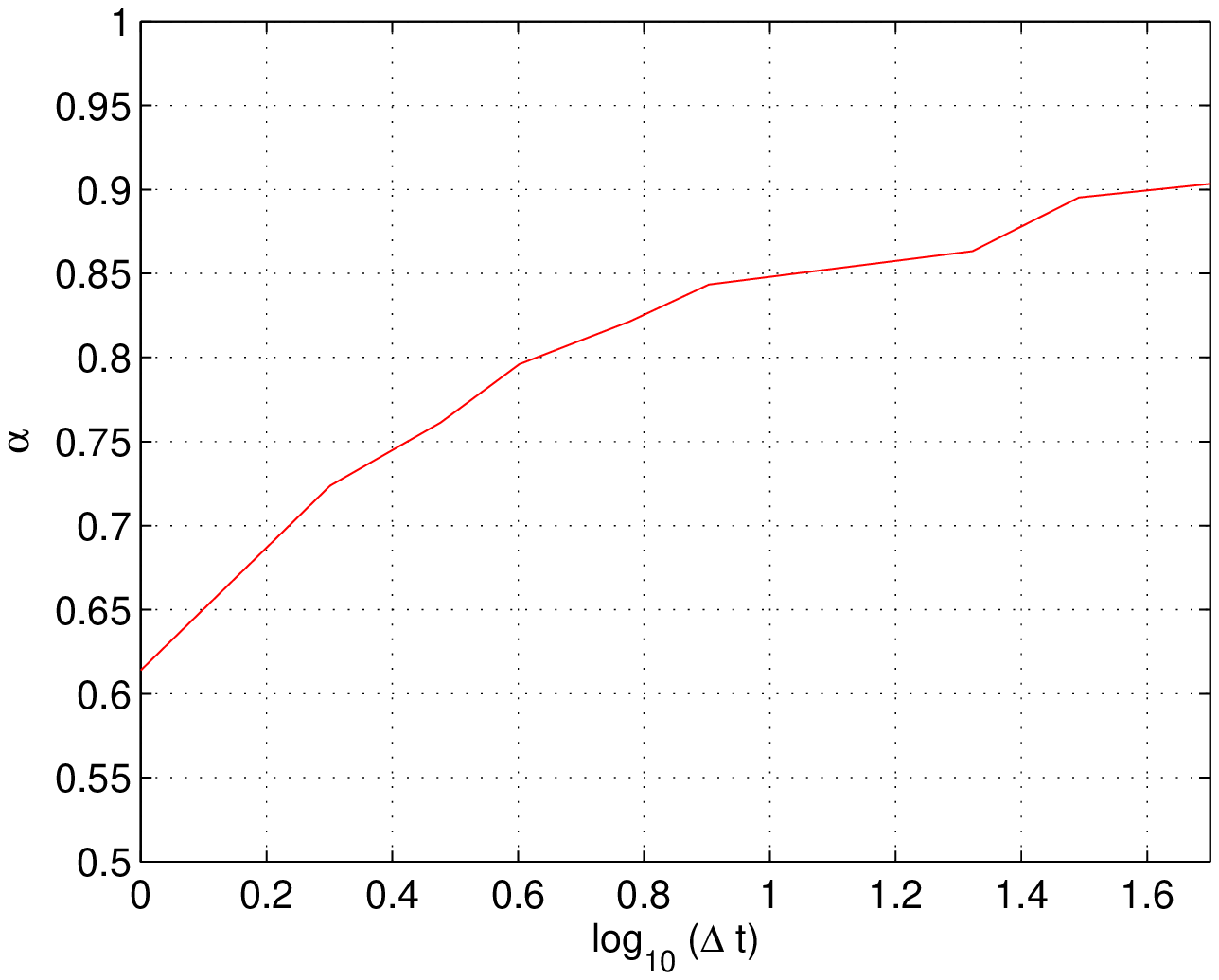}}
\centerline{\footnotesize
\begin{tabular}{p{7.5cm}} \bf Fig.\,5. \rm The transition behavior of $\alpha$ w.r.t the sampling time
  window produced by simulation.\end{tabular}}

\vskip 2mm

In the second test, the corresponding parameters of the model are
set as $T=1000000$, $\tau=1$. $N=20$, $k = [10, ..., 20]$. $M=2$,
$p_m = 1/2$ for $m=1$, $2$, $\lambda_m = [0.8, 1.1]$. The sampling
window size $\Delta t$ is controlled to vary from $1$ to $15$. As
shown in Fig.~5, $\alpha$ increases as $\Delta t$ is increased.


\vskip 2mm

\centerline{\includegraphics[width=3in]{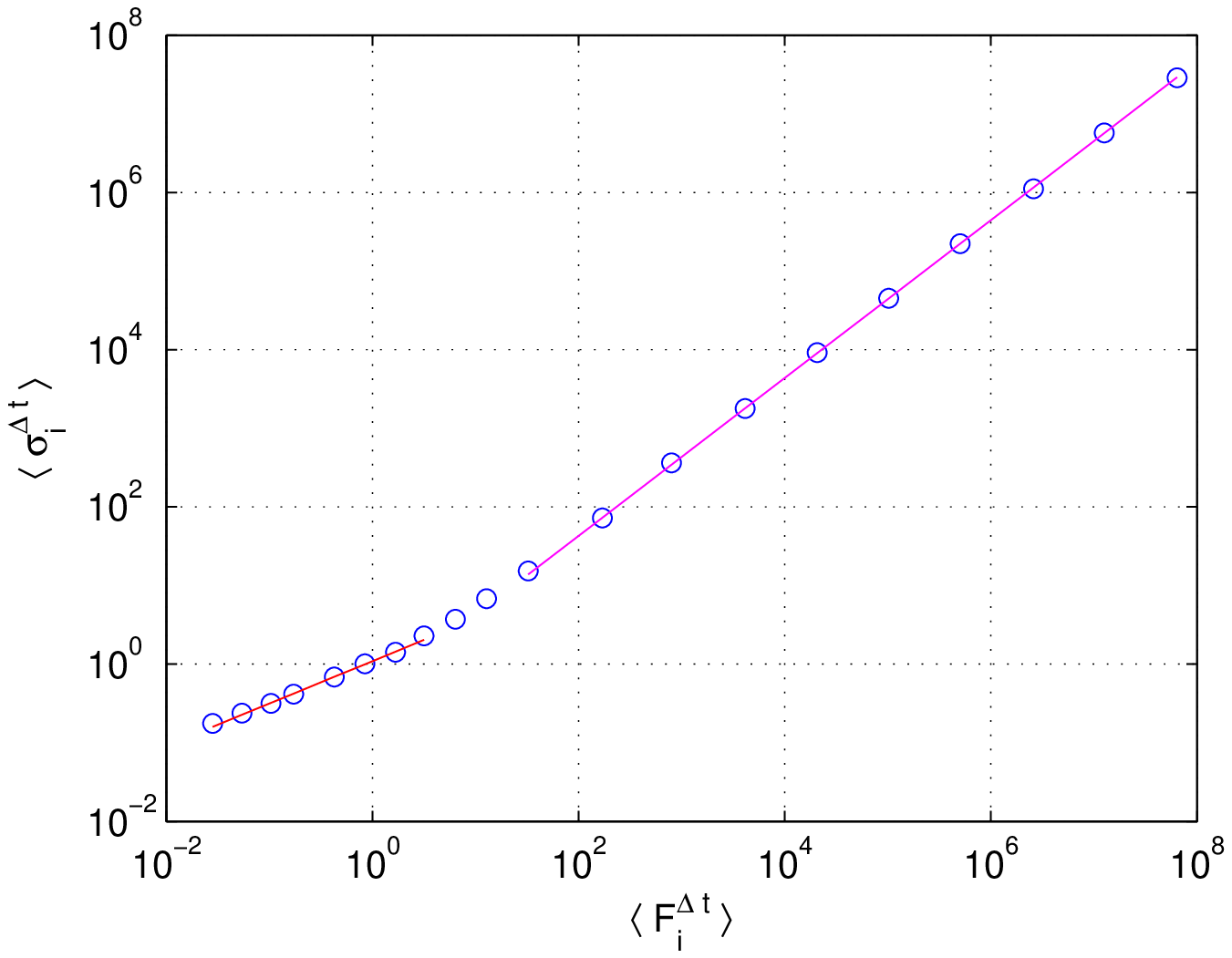}}
\centerline{\footnotesize
\begin{tabular}{p{7.5cm}} \bf Fig.\,6. \rm The crossover behavior produced by simulation.\end{tabular}}

\vskip 2mm

To test the existence of crossover behavior, the corresponding
parameters of the model are set as $T=100000$, $\tau=1$, $\Delta t
=1$. $M=10$, $p_m = 1/10$ for $m=1$, ..., $10$, $\lambda_m = [1, 2,
3, ..., 10]$. $N=20$, $k = [0.5, 0.5^2, ..., 0.5^8, 1, 2, 5, 5^2,
..., 5^{10}]$ in the third test. The result is plotted in Fig.~6,
which shows clear crossover behavior.


Another important phenomenon numerically reproduced by our model is
the multi-scaling behavior, which is the extension of the original
scaling law to the $q$th-order central moments
\begin{equation}
\label{eq.13}
  \sigma_{i,q} = \langle | F_i^{\Delta t} - \langle F_i^{\Delta t} \rangle|^q \rangle
  \propto \langle F_i^{\Delta t} \rangle ^{q \alpha (q)}
\end{equation}

Generally, this multi-scaling law (i.e. a dependence of $\alpha(q)$
on $q$) is assumed to be related with the multi-fractality of time
series \textsuperscript{\cite{EislerKertesz2005}}. In the fourth
test, the corresponding parameters of the model are set as
$T=10000$, $\tau=10$. $M=10$, $p_j = 1/10$ for $j=1$, ..., $10$,
$\lambda_j = [2, 3, ..., 11]$. $N=20$, $k = [1, 2, ..., 10]$. Fig.~7
plots the three $q$-$\alpha(q)$ plots for $\Delta t =1$, $\Delta t
=10$ and $\Delta t =100$, respectively. in which $\alpha(q)$ heavily
depends on $q$.

\vskip 2mm

\centerline{\includegraphics[width=3in]{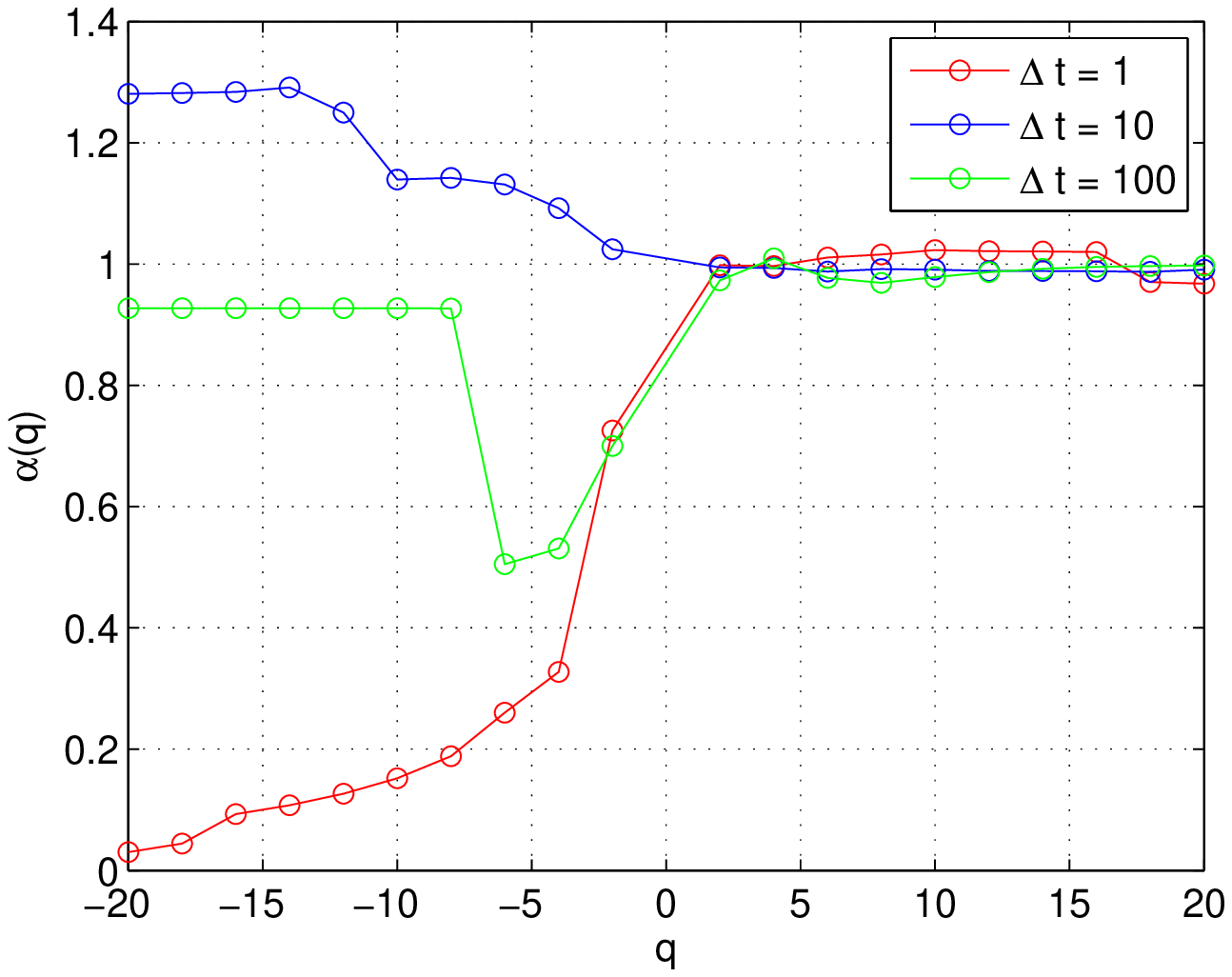}}
\centerline{\footnotesize
\begin{tabular}{p{7.5cm}} \bf Fig.\,7. \rm The multi-scaling behavior produced by simulation.\end{tabular}}

\vskip 2mm

\section{Conclusion}
\label{sec:conclusion}

In summary, a non-stationary Poisson process model is introduced in
this paper to explain the scaling of fluctuations in complex
systems. The scaling exponent $\alpha$ with non-universality values
between 1/2 and 1 can be analytically derived. The influences of the
sampling time window and the external/internal force ratio, the
crossover behavior, the multi-scaling phenomenon, and the connection
with long range dependency are also naturally explained. The model
is verified by numerical simulations.

Our work also sheds light on the well-known debate on whether
Poisson process models are applicable in modeling complex network
systems which often exhibit burstiness and long range dependency
\textsuperscript{\cite{PaxsonFloyd1995},
\cite{CaoWilliamLinDon2001}, \cite{KaragiannisMolleFaloutsos2004}}.
The results presented in this paper support the argument that the
non-stationary Poisson process are still powerful in modeling the
multi-scale behaviors of complex time-variant systems with
complicated interactions between external and internal dynamics.

\end{document}